\documentclass[showpacs,prl,twocolumn,aps,preprintnumbers,letterpaper,10pt]{revtex4}
\usepackage{epsfig}
\usepackage{epstopdf}
\usepackage{amsfonts,amsmath,amssymb,color, dsfont}
\usepackage{hyperref,bm,comment,graphicx}
\usepackage{natbib}

\def\slashchar#1{\setbox0=\hbox{$#1$}     		
   \dimen0=\wd0                                 	
   \setbox1=\hbox{/} \dimen1=\wd1               	
   \ifdim\dimen0>\dimen1                        	
      \rlap{\hbox to \dimen0{\hfil/\hfil}}      	
      #1                                        	
   \else                                        	
      \rlap{\hbox to \dimen1{\hfil$#1$\hfil}}   	
      /                                         	
   \fi}

\newcommand{\avg}[1]{\left \langle #1 \right \rangle}

\newcommand{\be}{\begin{equation}}
\newcommand{\ee}{\end{equation}}
\newcommand{\bear}{\begin{eqnarray}}
\newcommand{\eear}{\end{eqnarray}}
\newcommand{\ba}{\begin{array}}
\newcommand{\ea}{\end{array}}

\begin{document}

\title{Partial restoration of chiral symmetry in a confining string}

\author{Dmitri E. Kharzeev}
\affiliation{Department of Physics and Astronomy, Stony Brook University, Stony Brook, New York 11794-3800, USA}
\affiliation{Department of Physics,
Brookhaven National Laboratory, Upton, New York 11973-5000, USA}

\author{Frash\"er Loshaj}
\affiliation{Department of Physics and Astronomy, Stony Brook University, Stony Brook, New York 11794-3800, USA}

\date{\today}

\begin{abstract}
We attempt to describe the interplay of confinement and chiral symmetry breaking in QCD by using the string model. 
We argue that in the quasi-abelian picture of confinement based on the condensation of magnetic monopoles and the dual Meissner effect, the worldsheet dynamics of the confining string can be effectively described by the $1+1$ dimensional massless electrodynamics, which is exactly soluble. The transverse plane distribution of the chromoelectric field stretched between the quark and antiquark sources can then be attributed to the fluctuations in the position of the string. The dependence of the chiral condensate in the string on the (chromo-)electric field can be evaluated analytically, and is determined by the chiral anomaly and the $\theta$-vacuum structure. Therefore, our picture allows to predict the distribution of the chiral condensate in the plane transverse to the axis connecting the quark and antiquark. This prediction is compared to the lattice QCD results; a good agreement is found. 
 \end{abstract}

\pacs{
}

\maketitle

The interplay of confinement and chiral symmetry breaking in QCD is among the most prominent unsolved problems in modern physics. In fact,  none of the two phenomena have been understood  from the first principles so far. It thus may be useful to address a much more modest 
problem of understanding how the chiral symmetry breaking can occur once the presence of a confining background is {\it assumed}; this will be the subject of our paper.
\vskip0.3cm
We consider the mechanism of confinement based on the dual Meissner effect, as originally proposed by Nambu and Mandelstam \cite{tHooft,Mandelstam}. Seiberg and Witten \cite{Seiberg:1994rs} have demonstrated that this scenario is realized in ${\cal N}=2$ supersymmetric Yang-Mills theory, where magnetic monopoles at strong coupling become massless and condense. The resulting low-energy effective theory takes the form of an abelian gauge theory with matter. A small deformation of the ${\cal N}=2$ theory down to ${\cal N}=1$ then leads to the emergence of the abelian string that confines electric charges 
\cite{Seiberg:1994rs}. 
\vskip0.3cm

The confinement mechanism in QCD  may be different and can involve non-abelian strings (for a recent review, see  \cite{Shifman:2014jba}).
Nevertheless, the quasi-abelian picture based on the dual Meissner effect is supported by a number of lattice QCD studies, e.g. \cite{Singh:1993jj,Schilling:1998gz,Chernodub:2000rg,Chernodub:2005gz,Cea:2012qw}. 
  Most (but not all -- see \cite{Suzuki:2009xy}) of these studies rely on the use of a specific gauge -- so-called ``maximal abelian projection". 
  Let us summarize the arguments justifying this quasi-abelian approach. 
  \vskip0.3cm
  The basic observation is that the symmetry group $SU(N)$ of non-abelian gauge theories is compact, i.e. it is a topological group with a compact topology (for example, a sphere for $SU(2)$). The perturbation theory ignores this underlying topology, and this is a source of the disconnect that currently exists 
  between the perturbative and non-perturbative approaches to QCD \cite{Polyakov:1976fu,'tHooft:1981ht}. 
  The gauge fixing condition is at the root of the problem, since in general it leads to the emergence of unphysical propagating modes \cite{Gribov:1977wm,'tHooft:1981ht}. These unphysical modes can be removed in so-called "non-propagating" or "unitary"  gauges; 
  this leads to the emergence of singularities in space-time, which have a physical meaning \cite{'tHooft:1981ht}.  For example, in the abelian Higgs theory, this singularity describes a string-like magnetic vortex; the dynamics along the vortex line is effectively $(1+1)$ dimensional. The topological structure of the vortex is a consequence of the compact topology of the $U(1)$ gauge group.
  \vskip0.3cm
  For nonabelian gauge theories, the approach proposed in \cite{'tHooft:1981ht} is the following: first, fix the nonabelian part of the gauge redundancy by 
  reducing the gauge symmetry $SU(N)$ to that of the maximal abelian subgroup $U(1)^{N-1}$; we thus get the theory with $N-1$ different kinds of electric charges. The non-propagating gauge condition leads to point-like singularities in 3D space that are interpreted as magnetic monopoles with respect to 
  $U(1)^{N-1}$. The compactness of $U(1)^{N-1} \subset SU(N)$ is essential for this argument; for example in QCD the magnetic monopoles realize the map $\pi_2 (SU(3)/[U(1)]^2) = Z_2$. If these magnetic monopoles condense in the vacuum, they repel the electric fields and lead to the confinement of electric charges by means of the dual Meissner effect. The electric-magnetic duality plays an important role here; indeed, it is manifest in the Seiberg-Witten approach \cite{Seiberg:1994rs}. The point-like singularities emerge only when the maximal abelian subgroup $U(1)^{N-1}$ is unbroken; this is called the ``abelian projection" of $SU(N)$ \cite{'tHooft:1981ht}.
   \vskip0.3cm
Assuming this scenario for QCD, the magnetic monopoles condense in the vacuum, leading to the emergence of confining abelian electric flux tubes connecting quarks and antiquarks. These flux tubes are dual analogs of Abrikosov-Nilesen-Olesen (ANO) vortices in type II superconductors. 
It is well known that when a theory possesses charged massless chiral fermions, they can be localized within the cores of ANO vortices that contain magnetic field. The dynamics of these localized fermions is described by an effective $(1+1)$ dimensional theory, see e.g. \cite{Witten:1984eb}  -- the fermions can freely propagate along the vortex but cannot escape into the bulk of a superconductor.
In the dual Meissner picture of confinement, the QCD string is a dual ANO vertex containing the abelian electric field. 
\vskip0.3cm
Just like in the original ANO vertex, the core of a 
confining QCD string can contain localized charged chiral fermions, and their dynamics is described by an effective $(1+1)$ dimensional theory.  The 
abelian gauge  theory with massless fermions in $(1+1)$ dimensions is well known -- it is the Schwinger model with the lagrangean given by
\begin{equation}
\mathcal{L}=-\frac{1}{4}F_{\mu\nu}F^{\mu\nu}+\bar{\psi}(i\gamma^\mu \partial_\mu-g \gamma^\mu A_\mu)\psi ,
\label{eq:lag}
\end{equation}
where $g$ is the coupling constant with the dimension of mass. The theory is exactly soluble and possesses 
confinement, chiral symmetry breaking, axial anomaly, and the periodic $\theta$-vacuum \cite{Schwinger:1962tp,Lowenstein:1971fc,Coleman:1975pw}. 
In particular, it allows a computation of the chiral condensate as a function of electric field \cite{Hamer:1982mx}. Namely, the presence of the background electric field suppresses the magnitude of the chiral condensate, with a periodic dependence that originates from the $\theta$-vacuum of the theory. 
\vskip0.3cm
The string breaking in Schwinger model results from the periodic $\theta$ dependence, since the electric field  in $(1+1)$ dimensions plays the role of $\theta = 2\pi E/g$ angle, creating the imbalance between the left- and right-moving chiral fermions \cite{Coleman:1975pw}. It is tempting to speculate that in full $(3+1)$ dual Meissner picture the image of this string breaking process involves the Witten effect \cite{Witten:1979ey}: magnetic monopoles at finite  $\theta$ 
acquire an electric charge, which allows them to screen the confining potential -- see \cite{Gorsky:2011hd} for a specific realization of this scenario for nonabelian strings in dense QCD. The resulting dyon would thus behave as a fermion (or as a bosonic kink in the bosonized description of the Schwinger model), and may play the role of the produced quark.
\vskip0.3cm

Long time ago it was proposed to consider the Schwinger model as an effective theory of non-perturbative string fragmentation \cite{Casher:1974cks}, see also \cite{Fujita:1989vv,Wong:1991ub,Wong:2014ila}. Recently, we have revisited this approach by finding an explicit exact solution for the theory coupled to external fast quark sources \cite{Loshaj:2011jx,Kharzeev:2012re}. We found that the model provides a reasonably good description of the data on jet fragmentation in $e^+e^-$ annihilation, multiparticle production in $pp$ collisions, and the jet quenching in quark-gluon plasma. 
\vskip0.3cm
We have also identified the phenomenon of coherent coupled oscillations of chiral and vector charges in this theory \cite{Kharzeev:2013wra}, which can be viewed as a $(1+1)$ analog of the ``chiral magnetic wave" \cite{Kharzeev:2010gd,Burnier:2011bf} in 
$(3+1)$ dimensions -- indeed, both excitations are driven by the chiral anomaly, and the chiral magnetic wave is described by a $(1+1)$ theory in the limit of a strong magnetic field. We argued that this coherent oscillation of vector (electric) charge acts as an intense source of soft photon production, and may explain the ubiquitous enhancement of soft photons observed in hadronic processes \cite{Kharzeev:2013wra}. 

\vskip0.3cm

\begin{figure}[htbp]
\centering
\includegraphics[width=.8\linewidth]{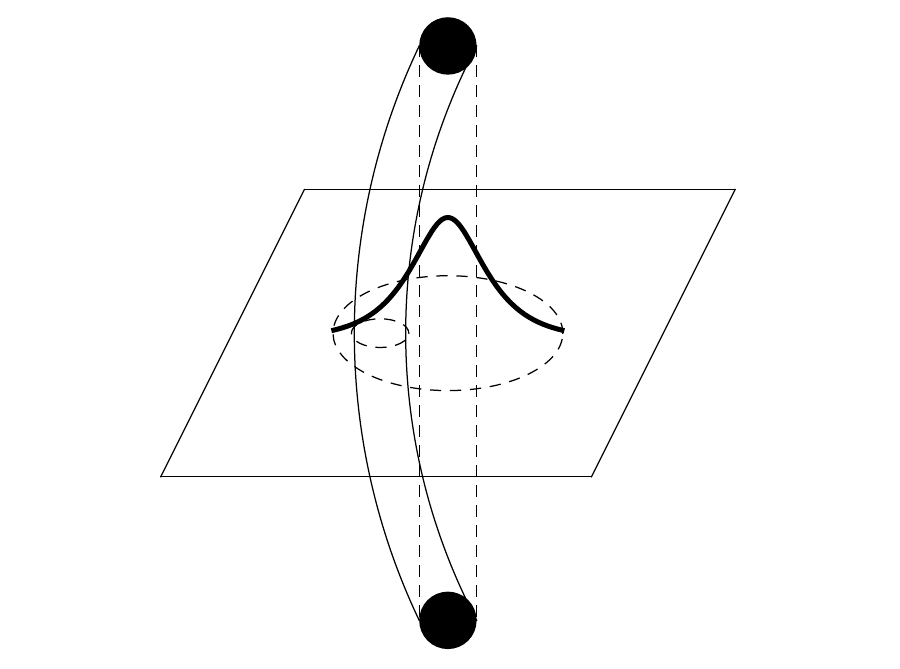} 
\caption{Thin string fluctuating in the transverse plane.}
\label{fig:StringFluctuation}
\end{figure}

In lattice QCD studies, the observation of the electric flux tube between color charges presents a clear indication of confinement. It has been observed that the transverse profile of the electric field resembles that of ANO vortex \cite{Schilling:1998gz,Cea:2012qw}. Recently a measurement of the chiral condensate in the presence of static charge was performed \cite{Iritani:2013rla}. The chiral condensate was found to be suppressed around the string, which indicates a partial restoration of chiral symmetry in the confining background. 
\vskip0.3cm
We will now argue that this partial restoration of chiral symmetry can be described if one assumes the presence of a ``thin" $(1+1)$ dimensional string with a position fluctuating in the transverse plane, as shown in Fig. \ref{fig:StringFluctuation}. 
\vskip0.3cm
Let us denote the physical $(3+1)$ dimensional electric field measured on the lattice by $E^{3+1}_{phys}(x_t)$ ($x_t$ is the coordinate in the plane transverse to the flux) and the electric field along the thin string in the $(1+1)$ description by $E^{1+1}$. We assume that both descriptions of the string should yield the same string tension, so the energy per unit length of the string should be equal:
\begin{eqnarray}\label{match}
\frac{1}{2}\int{d^2 x_t (E^{3+1}_{phys}(x_t))^2}=\frac{1}{2}(E^{1+1})^2
\label{eq:str}
\end{eqnarray}
Let us now introduce $P(x_t)$ as the probability for finding the thin string in the transverse plane at a position $x_t$. 
From the condition (\ref{match}), we can constrain this probability distribution: 
\begin{equation}
(E^{3+1}_{phys}(x_t))^2=(E^{1+1})^2 P(x_t).
\label{eq:pd}
\end{equation}
Since
\begin{equation}
\int{d^2 x_t P(x_t)}=1,
\label{eq:norm}
\end{equation}
the requirement \eqref{eq:str} is fulfilled. 
\vskip0.3cm

Now we can use the lattice data on the distribution of electric field to extract the probability distribution \eqref{eq:pd}. The knowledge of the dependence of the chiral condensate in the $1+1$ dimensional theory on the electric field together with the probability distribution will then allow us to predict the distribution 
of the chiral condensate around the confining flux tube. 
\vskip0.3cm

The Schwinger model is exactly soluble is by bosonization \cite{Lowenstein:1971fc,Coleman:1975pw}:
\begin{eqnarray}
\bar{\psi}\gamma^\mu\partial_\mu\psi &\rightarrow& \frac{1}{2}\partial_\mu \phi \partial^\mu \phi \nonumber \\
\bar{\psi}\gamma^\mu\psi &\rightarrow& -\frac{1}{\sqrt{\pi}}\epsilon^{\mu\nu}\partial_\nu\phi \nonumber \\
\bar{\psi}\gamma^\mu\gamma^5\psi &\rightarrow& \frac{1}{\sqrt{\pi}}\partial^\mu\phi,
\label{eq:bos}
\end{eqnarray}
Using these relations in the original Lagrangian \eqref{eq:lag}, it can be shown that the original massless fermionic theory is equivalent to the theory of a free massive scalar
\begin{equation}
\mathcal{L}=\frac{1}{2}\partial_\mu \phi \partial^\mu \phi-\frac{1}{2}\frac{g^2}{\pi}\phi^2.
\label{eq:lagsc}
\end{equation}
The expression for the chiral condensate in terms of the scalar field is 
\begin{equation}
\bar{\psi}\psi=-\frac{g e^\gamma}{2\pi^{3/2}}\cos(2\sqrt{\pi}\phi),
\label{eq:}
\end{equation}
where $\gamma\approx 0.5772$ is the Euler number. 
\vskip0.3cm

The chiral condensate can be evaluated through the Feynman-Hellmann theorem by differentiating the energy of the ground state in the presence of an electric field $E^{1+1}$ with respect to the fermion mass $m$, in the chiral limit $m=0$ \cite{Hamer:1982mx} :
\begin{equation}
\avg{\bar{\psi}\psi(x)}_{E^{1+1}}=-\frac{g e^\gamma}{2\pi^{3/2}}\cos\left(\frac{2\pi E^{1+1}}{g}\right).
\label{eq:conden}
\end{equation}
where $x$ is the longitudinal coordinate. We see that the value of the condensate is constant along the string and  depends only on the value of the background electric field. 
\vskip0.3cm

Let us now assume that the thin string fluctuates in the transverse plane (see Fig. 1), and the corresponding probability distribution is $P(x_t)$ normalized  by (\ref{eq:norm}). If the effective radius of the string is $a$, then the probability to find a 
string at a given transverse position is given by the integral of $P(x_t)$ over the string area, i.e. $\pi a^2 P(x_t)$. 
If the string with its electric field is present at a given $x_t$, it will modify the value of the chiral condensate according to 
(\ref{eq:conden}). If not, then there will be no electric field and the chiral condensate will not be modified, so within the Schwinger model it would be given by 
$$
\avg{\bar{\psi}\psi} (E^{1+1} =0) \equiv \avg{\bar{\psi}\psi}_{0}.
$$
Therefore in this picture the chiral condensate in the transverse plane can be computed as
\begin{eqnarray}
\avg{\bar{\psi}\psi(x_t)}&=&\left(1-\pi a^2 P(x_t)\right)\avg{\bar{\psi}\psi}_{0} \nonumber \\
&+&\pi a^2 P(x_t)\avg{\bar{\psi}\psi}_{E^{1+1}} .
\label{eq:probform}
\end{eqnarray}
We will see below that from the fit to the lattice data the value of the effective radius of the string $a$ appears comparable to the lattice spacing, i.e. the string is indeed "thin". 
\vskip0.3cm

\begin{figure}[htbp]
\centering
\includegraphics[width=1\linewidth]{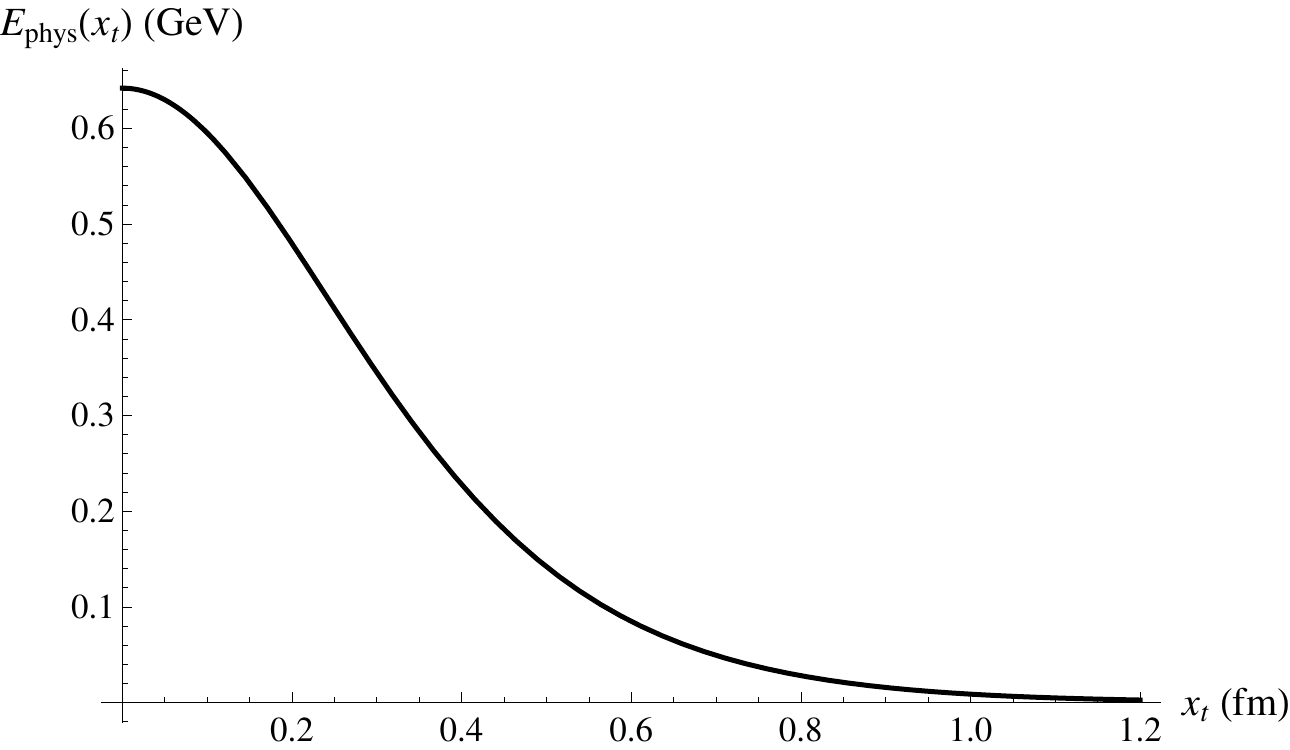} 
\caption{The longitudinal chromoelectric field between two static charges as given by \eqref{eq:efpar}.}
\label{fig:efield}
\end{figure}

\begin{figure}[htbp]
\centering
\includegraphics[width=.8\linewidth]{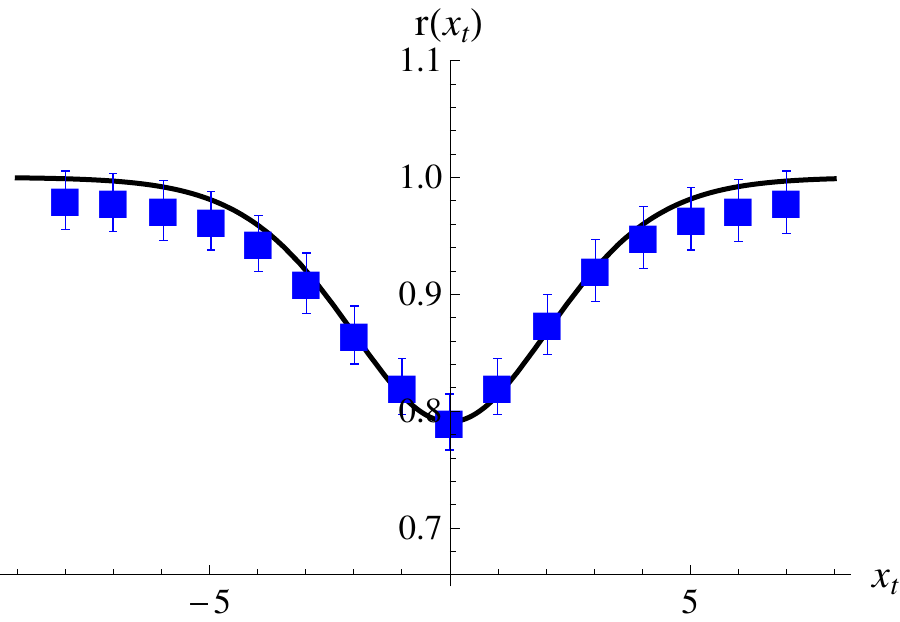} 
\caption{The chiral condensate around a confining flux tube as a function of the transverse distance. The solid line is the result of our model based on \eqref{eq:req}. The squares are from \cite{Iritani:2013rla}.}
\label{fig:condplot}
\end{figure}

In a recent lattice study \cite{Iritani:2013rla}, the authors compute on the lattice the following observable that quantifies the effect of confining flux tube on the chiral condensate:
\begin{equation}
r(x_t)=\frac{\avg{\bar{q}q(x_t) W}}{\avg{\bar{q}q}\avg{W}},
\label{eq:}
\end{equation} 
where $W$ is the Wilson loop operator of the static quarks. In our model, this quantity is given by
\begin{equation}
r(x_t)=\frac{\avg{\bar{\psi}\psi(x_t)}}{\avg{\bar{\psi}\psi}_{0}}
\label{eq:req}
\end{equation}
The suppression of the chiral condensate has been also described recently in terms of the $\sigma$ meson cloud surrounding the string  \cite{Kalaydzhyan:2014tfa}. 
\vskip0.3cm
To evaluate this quantity from (\ref{eq:probform}), we now need an independent information on the probability distribution  $P(x_t)$. Since $P(x_t)$ is the probability to find a longitudinal chromoelectric field at a given point $x_t$, the most direct source of information about it is the profile of chromoelectric field in the confining flux tube. 
There have been many lattice studies of the profile of the chromoelectric field between two static color charges. Here we use the recent lattice results of \cite{Cea:2012qw}. The measured chromoelectric field, as a function of the transverse coordinate, was shown to be described well by the following parameterization: 
\begin{equation}
E(x_t)=\frac{\phi}{2\pi}\frac{\mu^2}{\alpha}\frac{K_0[(\mu^2 x_t^2+\alpha^2)^{1/2}]}{K_1(\alpha)},
\label{eq:efpar}
\end{equation} 
where the values of the parameters above depend on the lattice coupling constant $\beta=2N/g^2$ and the number of "cooling steps" used to remove the short wavelength fluctuations. In \cite{Cea:2012qw}, the parameters $\mu,\phi$ and $\alpha$ were computed for four values of the coupling. In Fig. \ref{fig:efield}, we plot the profile of the electric field as a function of the transverse coordinate, computed at $\beta=6.01$ with 10 cooling steps.
We still have to fix the value of $a$ in order to compare with data. Using the values of parameters given above, we found that the value $a=1.12\ a_{latt}$, 
where $a_{latt}$ is the lattice spacing (which also depends on the coupling) provides a good description of the chiral condensate distribution in the vicinity of the flux tube, as shown in Fig. \ref{fig:condplot}. The effective radius of the string appears comparable to the lattice spacing, so it is indeed "thin". 
\vskip0.3cm
From Fig. 3, one can see that our simple model of fluctuating thin string describes the lattice results quite well. This lends additional support to the dual Meissner mechanism of confinement, and suggests that the longitudinal dynamics along the core of the string can be adequately described by the dimensionally reduced $(1+1)$ dimensional model. In the future, it would be interesting to extend this approach to nonabelian strings. 
\vskip0.3cm
Another promising direction is to apply our findings to the phenomenology of nonperturbative jet fragmentation. We have already observed that the longitudinal momentum distributions within a jet are adequately described by the $(1+1)$ string model \cite{Loshaj:2011jx,Kharzeev:2012re,Kharzeev:2013wra}. The Fourier transform of the transverse coordinate distribution of the "thin string" extracted from the lattice data in this paper may allow to describe also the nonperturbative transverse momentum distribution inside the jet; this introduces the "intrinsic" transverse momentum $k_t \sim 1$ GeV as required by the data. 
\section{Acknowledgements}

We thank L. Cosmai, A. Gorsky, T. Kalaydzhyan, and E. Shuryak for useful discussions. This work was
supported by the U.S. Department of Energy under Contracts No. DE-FG-88ER40388 and
DE-AC02-98CH10886. 


\begin{thebibliography}{99} \frenchspacing

\bibitem{tHooft}
  G. t' Hooft, in High Energy Physics, Ed. A. Zichichi, Edotorice Compositori,
Bologna (1975).

\bibitem{Mandelstam}
S. Mandelstam, Phys. Rep. \ {\bf 23C} (1976) 245.

\bibitem{Seiberg:1994rs} 
  N.~Seiberg and E.~Witten,
  Nucl.\ Phys.\ B {\bf 426}, 19 (1994)
  [Erratum-ibid.\ B {\bf 430}, 485 (1994)]
  [hep-th/9407087].

\bibitem{Shifman:2014jba} 
  M.~Shifman and A.~Yung,
  arXiv:1401.7067 [hep-th].
  
\bibitem{Singh:1993jj} 
  V.~Singh, D.~A.~Browne and R.~W.~Haymaker,
  Phys.\ Lett.\ B {\bf 306}, 115 (1993)
  [hep-lat/9301004].

\bibitem{Schilling:1998gz} 
  K.~Schilling, G.~S.~Bali and C.~Schlichter,
  Nucl.\ Phys.\ Proc.\ Suppl.\  {\bf 73}, 638 (1999)
  [hep-lat/9809039].
  
\bibitem{Chernodub:2000rg} 
  M.~N.~Chernodub, F.~V.~Gubarev, M.~I.~Polikarpov and V.~I.~Zakharov,
  Nucl.\ Phys.\ B {\bf 600}, 163 (2001)
  [hep-th/0010265].
  
\bibitem{Chernodub:2005gz} 
  M.~N.~Chernodub, K.~Ishiguro, Y.~Mori, Y.~Nakamura, M.~I.~Polikarpov, T.~Sekido, T.~Suzuki and V.~I.~Zakharov,
  Phys.\ Rev.\ D {\bf 72}, 074505 (2005)
  [hep-lat/0508004].
  
\bibitem{Cea:2012qw} 
  P.~Cea, L.~Cosmai and A.~Papa,
  Phys.\ Rev.\ D {\bf 86}, 054501 (2012)
  [arXiv:1208.1362 [hep-lat]].
  
\bibitem{Suzuki:2009xy} 
  T.~Suzuki, M.~Hasegawa, K.~Ishiguro, Y.~Koma and T.~Sekido,
  Phys.\ Rev.\ D {\bf 80}, 054504 (2009)
  [arXiv:0907.0583 [hep-lat]].
  
\bibitem{Polyakov:1976fu} 
  A.~M.~Polyakov,
  Nucl.\ Phys.\ B {\bf 120}, 429 (1977).
  
\bibitem{'tHooft:1981ht} 
  G.~'t Hooft,
  Nucl.\ Phys.\ B {\bf 190}, 455 (1981).
  
\bibitem{Gribov:1977wm} 
  V.~N.~Gribov,
  Nucl.\ Phys.\ B {\bf 139}, 1 (1978).
  
\bibitem{Witten:1984eb} 
  E.~Witten,
  Nucl.\ Phys.\ B {\bf 249}, 557 (1985).
  
\bibitem{Schwinger:1962tp}
  J.~S.~Schwinger,
  Phys.\ Rev.\  {\bf 128}, 2425-2429 (1962).

\bibitem{Lowenstein:1971fc}
  J.~H.~Lowenstein, J.~A.~Swieca,
  Annals Phys.\  {\bf 68}, 172-195 (1971).
  
\bibitem{Coleman:1975pw}
  S.~R.~Coleman, R.~Jackiw, L.~Susskind,
  Annals Phys.\  {\bf 93}, 267 (1975).

  
  
\bibitem{Hamer:1982mx} 
  C.~J.~Hamer, J.~B.~Kogut, D.~P.~Crewther and M.~M.~Mazzolini,
  Nucl.\ Phys.\ B {\bf 208}, 413 (1982).
  
   
  
\bibitem{Witten:1979ey} 
  E.~Witten,
  Phys.\ Lett.\ B {\bf 86}, 283 (1979).
  
\bibitem{Gorsky:2011hd} 
  A.~Gorsky, M.~Shifman and A.~Yung,
  Phys.\ Rev.\ D {\bf 83}, 085027 (2011)
  [arXiv:1101.1120 [hep-ph]].
  
\bibitem{Casher:1974cks}
  A.~Casher and J. ~Kogut and L.~Susskind,
  Phys.\ Rev.\ D\  {\bf 10}, 732 (1974).

\bibitem{Fujita:1989vv}
  T.~Fujita, J.~Hufner,
  Phys.\ Rev.\  {\bf D40}, 604-606 (1989).

\bibitem{Wong:1991ub}
  C.~Y.~Wong, R.~- C.~Wang, C.~C.~Shih,
  Phys.\ Rev.\  {\bf D44}, 257-262 (1991); 
  J.~Liu, C.~Y.~Wong, C.~C.~Shih, R.~-C.~Wang,
  Phys.\ Lett.\  {\bf B326}, 154-160 (1994).
  
\bibitem{Wong:2014ila} 
  C.~-Y.~Wong,
  arXiv:1404.0040 [hep-ph].

\bibitem{Loshaj:2011jx} 
  F.~Loshaj and D.~E.~Kharzeev,
  Int.\ J.\ Mod.\ Phys.\ E {\bf 21}, 1250088 (2012)
  [arXiv:1111.0493 [hep-ph]].
  
\bibitem{Kharzeev:2012re} 
  D.~E.~Kharzeev and F.~Loshaj,
  Phys.\ Rev.\ D {\bf 87}, 077501 (2013)
  [arXiv:1212.5857 [hep-ph]].

\bibitem{Kharzeev:2013wra} 
  D.~E.~Kharzeev and F.~Loshaj,
  arXiv:1308.2716 [hep-ph]; Phys. Rev. D, in press.
  
\bibitem{Kharzeev:2010gd} 
  D.~E.~Kharzeev and H.~-U.~Yee,
  Phys.\ Rev.\ D {\bf 83}, 085007 (2011)
  [arXiv:1012.6026 [hep-th]].
 
\bibitem{Burnier:2011bf} 
  Y.~Burnier, D.~E.~Kharzeev, J.~Liao and H.~-U.~Yee,
  Phys.\ Rev.\ Lett.\  {\bf 107}, 052303 (2011)
  [arXiv:1103.1307 [hep-ph]].
  
\bibitem{Kalaydzhyan:2014tfa} 
  T.~Kalaydzhyan and E.~Shuryak,
  arXiv:1402.7363 [hep-ph];
  arXiv:1404.1888 [hep-ph].
  
\bibitem{Iritani:2013rla} 
  T.~Iritani, G.~Cossu and S.~Hashimoto,
  arXiv:1311.0218 [hep-lat].

\end{thebibliography}
\bibliographystyle{unsrt}

\end{document}